# High critical current densities in superconducting $MgB_2$ thin films

S. H. Moon[a)], J. H. Yun, H. N. Lee, J. I. Kye, H. G. Kim, W. Chung, and B. Oh

LG Electronics Institute of Technology, Seoul 137-724, Korea

Superconducting $MgB_2$ thin films were prepared on $Al_2O_3$(0001) and MgO(100) substrates. Boron thin films were deposited by the electron-beam evaporation followed by post-annealing process with magnesium. Proper post annealing conditions were investigated to grow good superconducting $MgB_2$ thin films. The X-ray diffraction patterns showed randomly orientated growth of $MgB_2$ phase in our thin films. The surface morphology was examined by scanning electron microscope (SEM) and atomic force microscope (AFM). Critical current density ($J_c$) measured by transport method was about $10^7$ A/cm$^2$ at 15 K, and superconducting transition temperature ($T_c$) was ~ 39 K in the $MgB_2$ thin films on $Al_2O_3$.

---

[a)] Corresponding author, e-mail: smoon@LG-Elite.com

The recent discovery of the superconductivity above 39 K in magnesium boride ($MgB_2$) material attracts many researchers in scientific as well as technical reasons.[1] This material seems to have conventional BCS type superconductivity, and it has the simple structure.[2] The possibility of making good superconducting $MgB_2$ wires with low-cost was reported by several groups, because the weak-link problem between grains does not seem to be the case in this material.[3-5] The possibility of the electronic device application is also wide open, since it becomes possible to operate devices made of $MgB_2$ thin films with a low-cost refrigerator because of the higher $T_c \sim 39$ K than other conventional superconductors. In addition, it may be easy to make very reliable electronic devices and Josephson junctions with this material, because of its simpler crystal structure and longer coherence length compared with the oxide superconductors. To make electronic devices, $MgB_2$ thin films with good superconducting properties are essential. Several groups have reported $MgB_2$ thin films made by pulsed laser deposition (PLD) method or e-beam evaporation followed by post annealing.[6-11]

In this paper, we report our results on the growth of $MgB_2$ thin films by the electron-beam evaporation method followed by post-annealing process. The evaporation method has an advantage to the PLD method to make large area thin films. We investigated optimum growth conditions to make good superconducting $MgB_2$ thin films on $Al_2O_3$(0001) and MgO(100) substrates. Some superconducting transport properties ($T_c$ and $J_c$) and surface morphology of the $MgB_2$ thin films were also investigated.

To make $MgB_2$ thin films, we have started with the boron thin films deposited on the substrates, similar to the $MgB_2$ wire formation by Canfield *et al*.[4] The boron thin film was deposited by the electron beam evaporation from boron source in crucible. We have used two different substrate temperatures for the boron deposition, room temperature and 750 °C. The background pressure of the deposition chamber was below $1 \times 10^{-6}$ Torr. Typically 250 ~ 300 nm thick boron films were made with the deposition rate of ~ 2 Å/sec. The boron films as deposited were insulating and had brown color, and the broad boron peak was observed in X-ray diffraction pattern.

To obtain superconducting $MgB_2$ phase, the boron films were annealed under magnesium vapor environment. In order to maintain sufficient magnesium vapor environment during annealing, the boron films together with magnesium pieces were wrapped in a tantalum foil. Several pieces of titanium were also wrapped together with them as a spacer to prevent the direct contact of the boron film to the tantalum foil. Then, this whole thing was encapsulated in a quartz tube. To find an optimum annealing condition, we have varied annealing temperature and annealing time. The thickness of thin films has increased about 70 ~ 80 % due to the annealing. Final $MgB_2$ thin film thickness was in the range of 450 ~ 500 nm, starting from 250 ~ 300 nm thick boron films.

Figure 1 shows the normalized resistance data as a function of temperature ($R$ vs. $T$) for the $MgB_2$ thin films on $Al_2O_3(0001)$ and $MgO(100)$ substrates annealed at 800 $^o$C for 30 minutes. The $MgB_2$ thin films on $Al_2O_3$ substrate has been made from two kinds of boron films, deposited at room temperature or 750 $^o$C. The $MgB_2$ thin film made from the room temperature deposited boron film has shown higher $T_c$ than the other one. We have obtained the superconducting transition temperature ($T_c$) of ~39 K with the transition width ($\Delta T_c$) of about 0.3 K in the $MgB_2$ thin film on $Al_2O_3$. The $MgB_2$ thin film on $Al_2O_3$, made from the boron film deposited at 750 $^o$C, has shown $T_c$ of ~ 37.4 K with $\Delta T_c$ of about 0.7 K. Resistivity of this sample is about 2 times larger than the previous one, and it has a small tail in the $R$ vs. $T$ curve. This implies that the less formation of a crystalline phase of boron as a starting material was the better to form $MgB_2$ phase during the post-annealing. The $MgB_2$ film on MgO substrate has $T_c$ of ~ 38 K and $\Delta T_c$ of about 0.5 K. $MgB_2$ thin films on both $Al_2O_3$ and MgO have shown very similar $R$ vs. $T$ results, and the resistivity was in the range of 20 ~ 30 $\mu\Omega$-cm at room temperature for both films.

Figure 2 shows the glancing angle (incident angle of 5 degrees with sample rotation) X-ray diffraction patterns of the $MgB_2$ thin films, which is basically the powder pattern of the thin films. Mainly $MgB_2$ phase was observed, and there was no significant amount of MgO or $MgB_4$ phases, which was often observed in thin films made by e-beam evaporation with $MgB_2$ pressed pellets.[9]

There was also no evidence of preferred orientation from theta-two-theta X-ray diffraction patterns which are not shown here, in which the main peak of $MgB_2$ was overlapped with the substrate peak near ~ 43 degrees. From the X-ray diffraction patterns, we found that our $MgB_2$ thin films have grown in random orientation on $Al_2O_3$(0001) and MgO(100) substrates. The $MgB_2$ thin film made from the boron film deposited at 750 °C on $Al_2O_3$ has shown the similar random orientation growth.

Figure 3(a) and 3(b) show the scanning electron microscope (SEM) and atomic force microscope (AFM) images of the $MgB_2$ thin films on $Al_2O_3$ substrate with $T_c$ ~ 39 K. The surface of the boron film as deposited at room temperature was very flat and smooth from the SEM image. But the surface morphology of the $MgB_2$ thin films formed by the post-annealing process became a little rough as in Fig. 3. This $MgB_2$ thin film has the root mean square (RMS) roughness of ~ 20 nm according to the AFM image. Similar behavior was observed in $MgB_2$ thin films on MgO substrates. We note that the thin film looks very dense and its grain size is smaller than about 100 nm.

Figure 4 shows the critical current density ($J_c$) as a function of temperature for several $MgB_2$ thin films measured by four-point probe method. The inset shows the optical microscope image of the stripline used for the transport $J_c$ measurement. The $J_c$ criteria were less than 0.1 μV which was the noise level in the current-voltage characteristics (*I-V* curve). This stripline of 2 μm x 20 μm size was formed by photo-lithographic patterning and argon ion-milling with the milling rate of ~ 30 nm/min. The electrical contact was made by aluminum wire bonded directly to the contact pads. The contact resistance was less than ~ 1 Ω at 10 K.

The thickness of the $MgB_2$ thin films for $J_c$ measurement was in the range of 440 ~ 490 nm, and the error of the thickness measurement is estimated to be less than 20 %. The two films on $Al_2O_3$ were annealed at 800 °C for 30 minutes, and the thin film on MgO was annealed at 825 °C for 20 minutes. The transport $J_c$ of $MgB_2$ thin films on $Al_2O_3$ and MgO was about $10^7$ A/cm$^2$ in the temperature range of 10 ~ 15 K, which is the highest $J_c$ at zero field reported so far. The reason of this large $J_c$ is not clear yet, but we speculate the pinning at the grain boundary or at the interface between thin film and substrate may be very strong in our films. The $T_c$'s and $J_c$'s of MgB$_2$ thin films for various annealing conditions are summarized in Table I. In this table, we note very similar results were obtained from the two different substrates of Al$_2$O$_3$(0001) and MgO(100). We have obtained $T_c$ of 38 ~ 39 K and $J_c$ of ~ $10^7$ A/cm$^2$ in temperature range of 10 ~ 15 K from MgB$_2$ thin films on two different substrates of Al$_2$O$_3$(0001) and MgO(100).

In summary, we have made 450 ~ 500 nm thick MgB$_2$ thin films on Al$_2$O$_3$(0001) and MgO(100) substrates by the electron beam evaporation of boron and the post-annealing process. They have grown in random orientation. MgB$_2$ thin films of $T_c$ of ~ 39 K with $\Delta T_c$ ~ 0.3 K and the critical current density of 1.1 x $10^7$ A/cm$^2$ at 15 K was obtained on Al$_2$O$_3$(0001) substrates.

**Acknowledgements**

The authors acknowledge Mr. H. H. Kim for SEM, and Dr. S. M. Lee, Dr. J. H. Ahn, and Dr. H. J. Lee for helpful discussions. This work was supported by the Korean Ministry of Science & Technology under the National Research Laboratory project.

**References**

[1] J. Nagamatsu N. Nakagawa, T. Muranaka, Y. Zenitani, and J. Akimitsu, Nature **410**, 63 (2001).

[2] S. L. Bud'ko, G. Lapertot, C. Petrovic, C. E. Cunningham, N. Anderson, and P. C. Canfield, Phys. Rev. Lett. **86**, 1877 (2001).

[3] D. C. Larbalestier, M. O. Rikel, L. D. Cooley, A. A. Polyanskii, J. Y. Jiang, S. Patnaik, X. Y. Cai, D. M. Feldmann, A. Gurevich, A. A. Squitieri, M. T. Naus, C. B. Eom, E. E. Hellstrom, R. J. Cava, K. A. Regan, N. Rogado, M. A. Hayward, T. He, J. S. Slusky, P. Khalifah, K. Inumaru, and M. Haas, Nature 410, 186 (2001).

[4] P. C. Canfield, D. K. Finnemore, S. L. Bud'ko, J. E. Ostenson, G. Lapertot, C. E. Cunningham, and C. Petrovic, Phys. Rev. Lett. **86**, 2423 (2001).

[5] M. Kambara, N. Hari Babu, E. S. Sadki, J. R. Cooper, H. Minami, D. A. Cardwell, A. M. Campbell, and I. H. Inoue, Supercond. Sci. Technol. **14**, L5 (2001).

[6] W. N. Kang, H. -J. Kim, E. -M. Choi, C. U. Jung, S. I. Lee, cond-mat/0103179 (2001).

[7] C. B. Eom, M. K. Lee, J. H. Choi, L. Belenky, S. Patnaik, A. A. Polyanskii, E. E. Hellstrom, D. C. Larbalestier, N. Rogado, K. A. Regan, M. A. Hayward, T. He, J. S. Slusky, K. Inumaru, M. K. Haas, and R. J. Cava, cond-mat/0103425 (2001) (submitted to Nature).

[8] H. M. Christen, H. Y. Zhai, C. Cantoni, M. Paranthaman, B. C. Sales, C. Rouleau, D. P. Norton, D. K. Christen, and D. H. Lowndes, submitted to Physica C (March 22, 2001).

[9] S. Moon, J. H. Yun, J. I. Kye, H. K. Kim, and B. Oh, presented at the APS Meeting, Seattle, Washington, http://www.aps.org/meet/MAR01/mgb2/talk3.html#talk56 (March 12, 2001).

[10] A. Brinkman, D. Mijatovic, G. Rijnders, V. Leca, H. J. H. Smilde, I. Oomen, A. A. Golubov, F. Roesthuis, S. Harkema, H. Hilgenkamp, D. H. A. Blank, and H. Rogalla, cond-mat/0103198 (2001).

[11] M. Paranthaman, C. Cantoni, H. Y. Zhai, H. M. Christen, T. Aytug, S. Sathyamurthy, E. D. Specht, J. R. Thompson, D. H. Lowndes, H. R. Kerchner, and D. K. Christen, cond-mat/0103569 (2001) (submitted to Appl. Phys. Lett.).

**Table I.** The effect of the annealing temperature and time on the superconducting transition temperature ($T_c$) and critical current ($J_c$) of MgB$_2$ thin films on Al$_2$O$_3$(0001) and MgO(100) substrates. The MgB$_2$ thin films were formed from the boron thin films deposited at either room temperature (R.T.) or 750 °C.

| Boron Deposition Temperature | Substrate | Annealing Temperature (°C) | Annealing Time (minutes) | $T_c$ (K) Onset | $T_c$ (K) Zero | $J_c$(15 K, 0T) (10$^7$ A/cm$^2$) |
|---|---|---|---|---|---|---|
| R. T. | Al$_2$O$_3$ | 700 | 30 | - | - | |
| R. T. | Al$_2$O$_3$ | 750 | 30 | 38.7 | 37.3 | |
| R. T. | Al$_2$O$_3$ | 775 | 30 | 38.7 | 38.4 | |
| R. T. | Al$_2$O$_3$ | 800 | 30 | 39.2 | 38.9 | 1.1 |
| R. T. | Al$_2$O$_3$ | 825 | 30 | 39.0 | 38.7 | |
| R. T. | Al$_2$O$_3$ | 850 | 30 | 39.0 | 38.7 | |
| R. T. | Al$_2$O$_3$ | 900 | 30 | 38.9 | 38.5 | |
| R. T. | Al$_2$O$_3$ | 950 | 30 | 38.6 | 37.3 | |
| R. T. | Al$_2$O$_3$ | 800 | 20 | 39.1 | 38.8 | |
| R. T. | Al$_2$O$_3$ | 800 | 60 | 39.1 | 38.8 | |
| R. T. | MgO | 800 | 30 | 38.5 | 38.0 | |
| R. T. | MgO | 825 | 30 | 38.6 | 38.1 | 1.0 |
| 750 °C | Al$_2$O$_3$ | 800 | 30 | 37.7 | 37.0 | 0.5 |
| 750 °C | MgO | 800 | 30 | 38.1 | 37.4 | |

**Figure captions:**

FIG. 1. The normalized resistance as a function of temperature of $MgB_2$ thin films on $Al_2O_3(0001)$ and MgO(100) substrates. The circle indicates the thin films were made from the boron film deposited at 750 °C, and the other two samples were made from the boron films deposited at room temperature. All three samples were annealed at 800 °C for 30 minutes as explained in the text.

FIG. 2. The glancing angle (thin film) X-ray diffraction patterns of three $MgB_2$ thin films. MgB2 films made from boron films deposited on $Al_2O_3(0001)$ (a) at RT and (b) at 750 °C followed by annealing at 800 °C for 30 minutes, and (c) deposited on MgO(100) at RT annealed at 825 °C for 20 minutes. We note mainly $MgB_2$ phase with small unknown phases.

FIG. 3. (a) SEM and (b) AFM images of the $MgB_2$ thin film with $T_c$ ~ 39 K on $Al_2O_3(0001)$ substrate.

FIG. 4. Transport critical current density ($J_c$) of $MgB_2$ thin films as a function of temperature. The inset shows the narrow stripline of 2 μm x 20 μm size used for $J_c$ measurement by the four-point probe method.

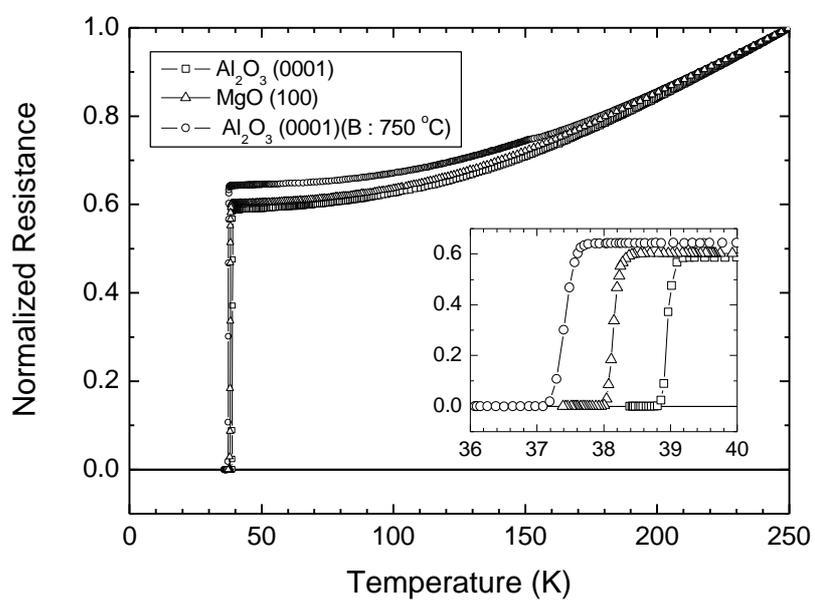

Fig. 1. S. H. Moon

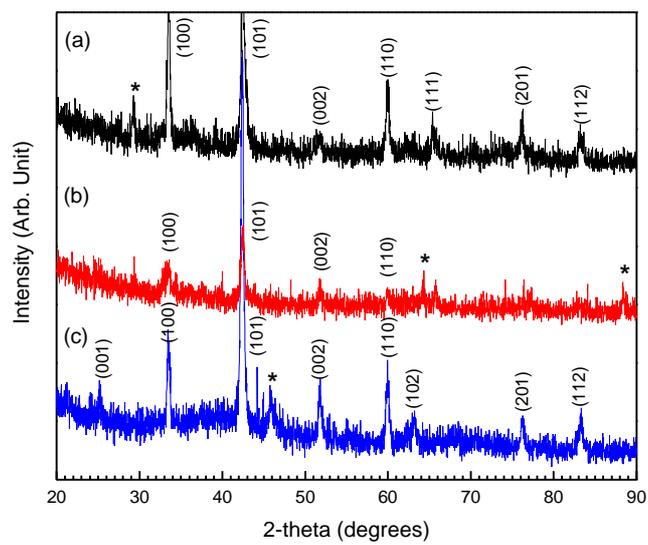

Fig. 2. S. H. Moon

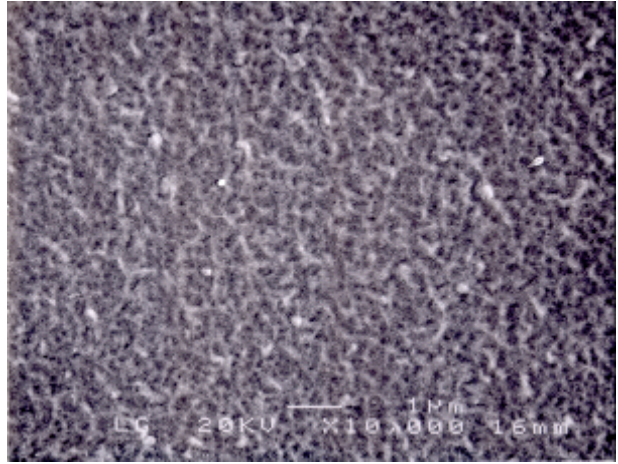

(a)

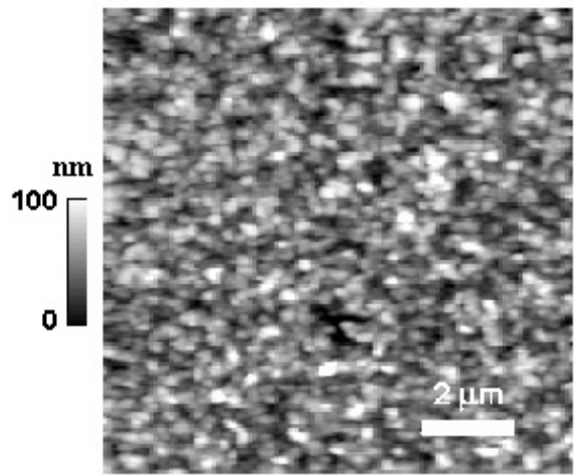

(b)

Fig. 3. S. H. Moon

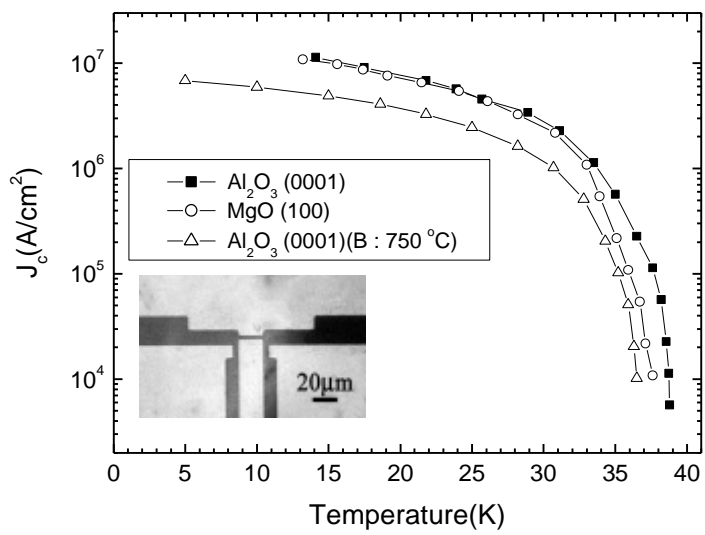

Fig. 4. S. H. Moon